\documentclass[preprint2]{aastex}
\usepackage{graphicx}
\graphicspath{{fig/}}
\begin{document}
%\preprint{NORDITA 2003-12 AP} Problems with aastex

% $Id: ms.tex,v 1.19 2003/10/22 08:03:53 brandenb Exp $
%|||||||||||||||||||||||||||||||||||||||||||||||||||||||||||||||||||
%             Customized Commands
%|||||||||||||||||||||||||||||||||||||||||||||||||||||||||||||||||||
%  mathematical abbreviations
%  =========================
%
% 12-feb-2003: wd introduced \BoldVec macro to get size of vectors in
%              exponents right
%
\newcommand{\BoldVec}[1]{\mathchoice%
  {\mbox{\boldmath $\displaystyle     #1$}}%
  {\mbox{\boldmath $\textstyle        #1$}}%
  {\mbox{\boldmath $\scriptstyle      #1$}}%
  {\mbox{\boldmath $\scriptscriptstyle#1$}}%
}
% math defs
\newcommand{\EQ}{\begin{equation}}
\newcommand{\EN}{\end{equation}}
\newcommand{\EQA}{\begin{eqnarray}}
\newcommand{\ENA}{\end{eqnarray}}
\newcommand{\eq}[1]{(\ref{#1})}
\newcommand{\EEq}[1]{Equation~(\ref{#1})}
\newcommand{\Eq}[1]{Eq.~(\ref{#1})}
\newcommand{\Eqs}[2]{Eqs~(\ref{#1}) and~(\ref{#2})}
\newcommand{\eqs}[2]{(\ref{#1}) and~(\ref{#2})}
\newcommand{\Eqss}[2]{Eqs~(\ref{#1})--(\ref{#2})}
\newcommand{\Sec}[1]{\S\,\ref{#1}}
\newcommand{\Secs}[2]{\S\S\,\ref{#1} and~\ref{#2}}
\newcommand{\Fig}[1]{Fig.~\ref{#1}}
\newcommand{\FFig}[1]{Figure~\ref{#1}}
\newcommand{\Tab}[1]{Table~\ref{#1}}
\newcommand{\Figs}[2]{Figures~\ref{#1} and \ref{#2}}
\newcommand{\Tabs}[2]{Tables~\ref{#1} and \ref{#2}}
\newcommand{\bra}[1]{\langle #1\rangle}
\newcommand{\bbra}[1]{\left\langle #1\right\rangle}
\newcommand{\mean}[1]{\overline #1}
\newcommand{\meanB}{\overline{B}}
\newcommand{\meanAA}{\overline{\mbox{\boldmath $A$}}}
\newcommand{\meanBB}{\overline{\mbox{\boldmath $B$}}}
\newcommand{\meanJJ}{\overline{\mbox{\boldmath $J$}}}
\newcommand{\meanuu}{\overline{\mbox{\boldmath $u$}}}
\newcommand{\meanAB}{\overline{\mbox{\boldmath $A\cdot B$}}}
\newcommand{\meanAoBo}{\overline{\mbox{\boldmath $A_0\cdot B_0$}}}
\newcommand{\meanApoBpo}{\overline{\mbox{\boldmath $A'_0\cdot B'_0$}}}
\newcommand{\meanApBp}{\overline{\mbox{\boldmath $A'\cdot B'$}}}
\newcommand{\meanuxB}{\overline{\mbox{\boldmath $\delta u\times \delta B$}}}
\newcommand{\mod}[1]{\mid\!\!#1\!\!\mid}
\newcommand{\chk}[1]{[{\em check: #1}]}
%\newcommand{\inst}[1]{$^{#1}$}
%
% tilde
%
\newcommand{\teps}{\tilde{\epsilon} {}}
%
%  unit vectors
%
\newcommand{\nnn}{\hat{\mbox{\boldmath $n$}} {}}
\newcommand{\vvv}{\hat{\mbox{\boldmath $v$}} {}}
\newcommand{\rr}{\hat{\mbox{\boldmath $r$}} {}}
\newcommand{\xxx}{\hat{\mbox{\boldmath $x$}} {}}
\newcommand{\yyy}{\hat{\mbox{\boldmath $y$}} {}}
\newcommand{\zz}{\hat{\mbox{\boldmath $z$}} {}}
\newcommand{\pp}{\hat{\mbox{\boldmath $\phi$}} {}}
\newcommand{\ttt}{\hat{\mbox{\boldmath $\theta$}} {}}
\newcommand{\OOO}{\hat{\mbox{\boldmath $\Omega$}} {}}
\newcommand{\ooo}{\hat{\mbox{\boldmath $\omega$}} {}}
\newcommand{\BBBB}{\hat{\mbox{\boldmath $B$}} {}}
%
%  vectors
%
\newcommand{\gggg}{\BoldVec{g} {}}
\newcommand{\ddd}{\BoldVec{d} {}}
\newcommand{\rrr}{\BoldVec{r} {}}
\newcommand{\xx}{\BoldVec{x}{}}
\newcommand{\yy}{\BoldVec{y} {}}
\newcommand{\zzz}{\BoldVec{z} {}}
\newcommand{\uu}{\BoldVec{u} {}}
\newcommand{\vv}{\BoldVec{v} {}}
\newcommand{\ww}{\BoldVec{w} {}}
\newcommand{\mm}{\BoldVec{m} {}}
\newcommand{\PP}{\BoldVec{P} {}}
\newcommand{\QQ}{\BoldVec{Q} {}}
\newcommand{\UU}{\BoldVec{U} {}}
\newcommand{\bb}{\BoldVec{b} {}}
\newcommand{\qq}{\BoldVec{q} {}}
\newcommand{\BB}{\BoldVec{B} {}}
\newcommand{\HH}{\BoldVec{H} {}}
\newcommand{\II}{\BoldVec{I} {}}
\newcommand{\AAA}{\BoldVec{A} {}}
\newcommand{\aaa}{\BoldVec{a} {}}
\newcommand{\aaaa}{\BoldVec{a} {}} %(convert aaa -> aaaa, compatibility problem)
\newcommand{\eee}{\BoldVec{e} {}}
\newcommand{\jj}{\BoldVec{j} {}}
\newcommand{\JJ}{\BoldVec{J} {}}
\newcommand{\nn}{\BoldVec{n} {}}
\newcommand{\ee}{\BoldVec{e} {}}
\newcommand{\ff}{\BoldVec{f} {}}
\newcommand{\EE}{\BoldVec{E} {}}
\newcommand{\FF}{\BoldVec{F} {}}
\newcommand{\TT}{\BoldVec{T} {}}
\newcommand{\CC}{\BoldVec{C} {}}
\newcommand{\KK}{\BoldVec{K} {}}
\newcommand{\MM}{\BoldVec{M} {}}
\newcommand{\GG}{\BoldVec{G} {}}
\newcommand{\kk}{\BoldVec{k} {}}
\newcommand{\SSS}{\BoldVec{S} {}}
\newcommand{\grav}{\BoldVec{g} {}}
\newcommand{\nab}{\BoldVec{\nabla} {}}
\newcommand{\OO}{\BoldVec{\Omega} {}}
\newcommand{\oo}{\BoldVec{\omega} {}}
\newcommand{\LL}{\BoldVec{\Lambda} {}}
\newcommand{\llambda}{\BoldVec{\lambda} {}}
\newcommand{\pomega}{\BoldVec{\varpi} {}}
%
%  correlation tensors
%
\newcommand{\SSSS}{\mbox{\boldmath ${\sf S}$} {}}
\newcommand{\BBB}{\mbox{\boldmath ${\cal B}$} {}}
\newcommand{\emf}{\mbox{\boldmath ${\cal E}$} {}}
\newcommand{\FFF}{\mbox{\boldmath ${\cal F}$} {}}
\newcommand{\GGG}{\mbox{\boldmath ${\cal G}$} {}}
\newcommand{\HHH}{\mbox{\boldmath ${\cal H}$} {}}
\newcommand{\QQQ}{\mbox{\boldmath ${\cal Q}$} {}}
\newcommand{\GGGG}{{\bf G} {}}
%
%  operators  (roman)
%
\newcommand{\grad}{{\rm grad} \, {}}
\newcommand{\curl}{{\rm curl} \, {}}
\newcommand{\dive}{{\rm div}  \, {}}
\newcommand{\Dive}{{\rm Div}  \, {}}
\newcommand{\DD}{{\rm D} {}}
\newcommand{\dd}{{\rm d} {}}
\newcommand{\const}{{\rm const}  {}}
\newcommand{\crit}{{\rm crit}  {}}
\def\degr{\hbox{$^\circ$}}
\def\la{\mathrel{\mathchoice {\vcenter{\offinterlineskip\halign{\hfil
$\displaystyle##$\hfil\cr<\cr\sim\cr}}}
{\vcenter{\offinterlineskip\halign{\hfil$\textstyle##$\hfil\cr<\cr\sim\cr}}}
{\vcenter{\offinterlineskip\halign{\hfil$\scriptstyle##$\hfil\cr<\cr\sim\cr}}}
{\vcenter{\offinterlineskip\halign{\hfil$\scriptscriptstyle##$\hfil\cr<\cr\sim\cr}}}}}
\def\ga{\mathrel{\mathchoice {\vcenter{\offinterlineskip\halign{\hfil
$\displaystyle##$\hfil\cr>\cr\sim\cr}}}
{\vcenter{\offinterlineskip\halign{\hfil$\textstyle##$\hfil\cr>\cr\sim\cr}}}
{\vcenter{\offinterlineskip\halign{\hfil$\scriptstyle##$\hfil\cr>\cr\sim\cr}}}
{\vcenter{\offinterlineskip\halign{\hfil$\scriptscriptstyle##$\hfil\cr>\cr\sim\cr}}}}}
%
%  numbers
%
\def\Ta{\mbox{\rm Ta}}
\def\Ra{\mbox{\rm Ra}}
\def\Ma{\mbox{\rm Ma}}
\def\Roo{\mbox{\rm Ro}^{-1}}
\def\Pra{\mbox{\rm Pr}}
\def\Pran{\mbox{\rm Pr}}
\def\Pm{\mbox{\rm Pr}_M}
\def\Rm{\mbox{\rm Re}_M}
\def\Rey{\mbox{\rm Re}}
\def\Pe{\mbox{\rm Pe}}
\newcommand{\ea}{{\em et al. }}
\newcommand{\eaa}{{\em et al. }}
\def\half{{\textstyle{1\over2}}}
\def\threehalf{{\textstyle{3\over2}}}
\def\onethird{{\textstyle{1\over3}}}
\def\twothird{{\textstyle{2\over3}}}
\def\fourthird{{\textstyle{4\over3}}}
\def\quarter{{\textstyle{1\over4}}}
\newcommand{\W}{\,{\rm W}}
\newcommand{\V}{\,{\rm V}}
\newcommand{\kV}{\,{\rm kV}}
\newcommand{\T}{\,{\rm T}}
\newcommand{\G}{\,{\rm G}}
\newcommand{\Hz}{\,{\rm Hz}}
\newcommand{\kHz}{\,{\rm kHz}}
\newcommand{\kG}{\,{\rm kG}}
\newcommand{\K}{\,{\rm K}}
\newcommand{\g}{\,{\rm g}}
\newcommand{\s}{\,{\rm s}}
\newcommand{\ms}{\,{\rm ms}}
\newcommand{\cm}{\,{\rm cm}}
\newcommand{\m}{\,{\rm m}}
\newcommand{\km}{\,{\rm km}}
\newcommand{\kms}{\,{\rm km/s}}
\newcommand{\kg}{\,{\rm kg}}
\newcommand{\Mm}{\,{\rm Mm}}
\newcommand{\pc}{\,{\rm pc}}
\newcommand{\kpc}{\,{\rm kpc}}
\newcommand{\yr}{\,{\rm yr}}
\newcommand{\Myr}{\,{\rm Myr}}
\newcommand{\Gyr}{\,{\rm Gyr}}
\newcommand{\erg}{\,{\rm erg}}
\newcommand{\mol}{\,{\rm mol}}
\newcommand{\dyn}{\,{\rm dyn}}
\newcommand{\J}{\,{\rm J}}
\newcommand{\RM}{\,{\rm RM}}
\newcommand{\EM}{\,{\rm EM}}
\newcommand{\AU}{\,{\rm AU}}
\newcommand{\A}{\,{\rm A}}
%
%  journals
%
\newcommand{\ycsf}[3]{ #1, {Chaos, Solitons \& Fractals,} {#2}, #3}
\newcommand{\yepl}[3]{ #1, {Europhys. Lett.,} {#2}, #3}
\newcommand{\yaj}[3]{ #1, {AJ,} {#2}, #3}
\newcommand{\yjgr}[3]{ #1, {JGR,} {#2}, #3}
\newcommand{\ysol}[3]{ #1, {Sol. Phys.,} {#2}, #3}
\newcommand{\yapj}[3]{ #1, {ApJ,} {#2}, #3}
\newcommand{\yapjl}[3]{ #1, {ApJ,} {#2}, #3}
\newcommand{\yapjs}[3]{ #1, {ApJ Suppl.,} {#2}, #3}
\newcommand{\yan}[3]{ #1, {AN,} {#2}, #3}
\newcommand{\ymhdn}[3]{ #1, {Magnetohydrodyn.} {#2}, #3}
\newcommand{\yana}[3]{ #1, {A\&A,} {#2}, #3}
\newcommand{\yanas}[3]{ #1, {A\&AS,} {#2}, #3}
\newcommand{\yanar}[3]{ #1, {A\&AR,} {#2}, #3}
\newcommand{\yass}[3]{ #1, {Ap\&SS,} {#2}, #3}
\newcommand{\ygafd}[3]{ #1, {Geophys. Astrophys. Fluid Dyn.,} {#2}, #3}
\newcommand{\ypasj}[3]{ #1, {Publ. Astron. Soc. Japan,} {#2}, #3}
\newcommand{\yjfm}[3]{ #1, {JFM,} {#2}, #3}
\newcommand{\ypf}[3]{ #1, {Phys. Fluids,} {#2}, #3}
\newcommand{\ypp}[3]{ #1, {Phys. Plasmas,} {#2}, #3}
\newcommand{\ysov}[3]{ #1, {Sov. Astron.,} {#2}, #3}
\newcommand{\yjetp}[3]{ #1, {Sov. Phys. JETP,} {#2}, #3}
\newcommand{\yphy}[3]{ #1, {Physica,} {#2}, #3}
\newcommand{\yannr}[3]{ #1, {ARA\&A,} {#2}, #3}
\newcommand{\yaraa}[3]{ #1, {ARA\&A,} {#2}, #3}
\newcommand{\yprs}[3]{ #1, {Proc. Roy. Soc. Lond.,} {#2}, #3}
\newcommand{\yprl}[3]{ #1, {PRL,} {#2}, #3}
\newcommand{\yphl}[3]{ #1, {Phys. Lett.,} {#2}, #3}
\newcommand{\yptrs}[3]{ #1, {Phil. Trans. Roy. Soc.,} {#2}, #3}
\newcommand{\ymn}[3]{ #1, {MNRAS,} {#2}, #3}
\newcommand{\ynat}[3]{ #1, {Nat,} {#2}, #3}
\newcommand{\ysci}[3]{ #1, {Sci,} {#2}, #3}
\newcommand{\ysph}[3]{ #1, {Solar Phys.,} {#2}, #3}
\newcommand{\ypr}[3]{ #1, {Phys. Rev.,} {#2}, #3}
\newcommand{\spr}[2]{ ~#1~ {\em Phys. Rev. }{\bf #2} (submitted)}
\newcommand{\ppr}[2]{ ~#1~ {\em Phys. Rev. }{\bf #2} (in press)}
\newcommand{\ypnas}[3]{ #1, {Proc. Nat. Acad. Sci.,} {#2}, #3}
\newcommand{\yicarus}[3]{ #1, {Icarus,} {#2}, #3}
\newcommand{\yspd}[3]{ #1, {Sov. Phys. Dokl.,} {#2}, #3}
\newcommand{\yjcp}[3]{ #1, {J. Comput. Phys.,} {#2}, #3}
\newcommand{\yjour}[4]{ #1, {#2}, {#3}, #4}
\newcommand{\yprep}[2]{ #1, {\sf #2}}
\newcommand{\ybook}[3]{ #1, {#2} (#3)}
\newcommand{\yproc}[5]{ #1, in {#3}, ed. #4 (#5), #2}
\newcommand{\pproc}[4]{ #1, in {#2}, ed. #3 (#4), (in press)}
\newcommand{\ppp}[1]{ #1, {Phys. Plasmas,} (in press)}
\newcommand{\sapj}[1]{ #1, {ApJ,} (submitted)}
\newcommand{\sana}[1]{ #1, {A\&A,} (submitted)}
\newcommand{\san}[1]{ #1, {AN,} (submitted)}
\newcommand{\sprl}[1]{ #1, {PRL,} (submitted)}
\newcommand{\pprl}[1]{ #1, {PRL,} (in press)}
\newcommand{\sjfm}[1]{ #1, {JFM,} (submitted)}
\newcommand{\sgafd}[1]{ #1, {Geophys. Astrophys. Fluid Dyn.,} (submitted)}
\newcommand{\pgafd}[1]{ #1, {Geophys. Astrophys. Fluid Dyn.,} (in press)}
\newcommand{\tana}[1]{ #1, {A\&A,} (to be submitted)}
\newcommand{\smn}[1]{ #1, {MNRAS,} (submitted)}
\newcommand{\pmn}[1]{ #1, {MNRAS,} (in press)}
\newcommand{\papj}[1]{ #1, {ApJ,} (in press)}
\newcommand{\papjl}[1]{ #1, {ApJL,} (in press)}
\newcommand{\sapjl}[1]{ #1, {ApJL,} (submitted)}
\newcommand{\pana}[1]{ #1, {A\&A,} (in press)}
\newcommand{\pan}[1]{ #1, {AN,} (in press)}
\newcommand{\pjour}[2]{ #1, {#2,} (in press)}

\title{Is nonhelical hydromagnetic turbulence peaked at small scales?}

\author{Nils Erland L.\ Haugen\altaffilmark{1},
Axel Brandenburg\altaffilmark{2},
Wolfgang Dobler\altaffilmark{3}}

\altaffiltext{1}{Department of Physics, The Norwegian University of Science
  and Technology, H{\o}yskoleringen 5, N-7034 Trondheim, Norway;
nils.haugen@phys.ntnu.no}
\altaffiltext{2}{NORDITA, Blegdamsvej 17, DK-2100 Copenhagen \O, Denmark;
brandenb@nordita.dk}
\altaffiltext{3}{Kiepenheuer-Institut f\"ur Sonnenphysik, 
Sch{\"o}n\-eck\-stra{\ss}e 6, D-79104 Freiburg, Germany; 
Dobler@kis.uni-freiburg.de}

\begin{abstract}
Nonhelical hydromagnetic turbulence without an imposed magnetic field
is considered in the case where the magnetic Prandtl number is unity.
The magnetic field is entirely due to dynamo action.
The magnetic energy spectrum peaks at a wavenumber of about 5 times the
minimum wavenumber in the domain, and not at the resistive scale, as
has previously been argued.
Throughout the inertial range the spectral magnetic energy exceeds
the kinetic energy by a factor of about 2.5, and both spectra are
approximately parallel.
At first glance, the total energy
spectrum seems to be close to $k^{-3/2}$, but there is a strong bottleneck effect
and it is suggested that the asymptotic spectrum is $k^{-5/3}$.
This is supported by the value of the second order structure function exponent
that is found to be $\zeta_2=0.70$, suggesting a $k^{-1.70}$ spectrum.
\end{abstract}

\section{Introduction}

It is generally accepted that in hydromagnetic turbulence the magnetic
field tends to be more intermittent than the velocity field.
This is evidenced by many numerical simulations where the magnetic field
is dynamo-generated \citep{MFP81,Kida,BJNRST,Kulsrud}.
Furthermore, linear theory predicts that the exponentially growing
magnetic energy spectrum increases with wavenumber like $k^{+3/2}$ \cite{Kaz68}.
Taken at face value, linear theory would suggest that the magnetic
energy spectrum should be peaked at the `resistive' cutoff scale.
In the case of the interstellar medium the resistive cutoff scale
would be tiny ($<10^{10}\cm$) in comparison with other relevant
scales ($>10^{18}\cm$).
There is of course no doubt that there are magnetic fluctuations in the
interstellar medium at a scale below $10^{10}\cm$, as evidenced by interstellar
scintillation measurements \citep{GS95}, but it remains implausible
that magnetic fields at such small scale contribute significantly
to the energy budget.

The question of small scale magnetic fields has worried theorists for
the last decade.
\citet{KA92} have shown that, independent of the possible presence of
magnetic helicity, the {\it kinematic} magnetic energy spectrum follows the Kazantsev
$k^{+3/2}$ law, and they speculate that this may suppress large scale dynamo
action (of $\alpha^2$ or $\alpha\Omega$ type); see also \citet{VC92}.
The case of helical dynamos is now reasonably well understood in the case
of large magnetic Reynolds number.
The saturation level of helical dynamos is not suppressed, but the
saturation time is the resistive time scale \citep{B01}.
This is not the result of small scale magnetic fields in general,
but due to the helical small scale fields that are produced by the
$\alpha$ effect as a by-product \citep{FB02,BB02}.

When the degree of kinetic helicity of the flow falls below a certain
threshold, no large scale dynamo action is possible and the magnetic
energy spectrum is peaked at a scale much smaller than the forcing
scale \citep{MB02}.
This may not be so much a concern for stellar dynamos where differential
rotation is important, but it could be a problem for dynamo action in the
interstellar medium and in clusters of galaxies.
Analytic approaches to the nonlinear saturation of nonhelical dynamos
have corroborated the notion that the magnetic energy spectrum peaks at
the resistive scale \citep{Scheko02a,Scheko02b}.
Numerical simulations have confirmed a peak well below the forcing scale
\citep{CV00a,MC01}, but a resolution of up to $256^3$ collocation points
is still insufficient to establish whether the location of the peak is
different from the resistive scale.

So far there has been no evidence for $k^{-5/3}$ inertial range
scaling of the magnetic energy as in the case of an imposed field
\citep{GS95,CV00b,MG01} or as in decaying hydromagnetic turbulence
\citep{BM00}.
A difficulty in establishing power law scaling is the
lack of a sufficiently long inertial range.
In some cases the use of hyper-resistivity (where the $\nabla^2$
diffusion operator is replaced for example by $-\nabla^4$, in order to
shorten the diffusive subrange) has led to the concern that
it may cause an artificially enhanced `bottleneck effect' with a shallower
$k^{-1}$ spectrum just before the dissipative subrange \citep{BM00}.
Yet another problem is the possibility of a physical bottleneck effect \citep{Fal94}
that will be more extreme in three-dimensional spectra than in the
one-dimensional spectra accessible from turbulence experiments
\citep{BM00,DHYB03}.

In this Letter we use simulations at resolutions of up to $1024^3$
meshpoints to study the form of the
magnetic and kinetic energy spectra in the inertial range in the case
where the turbulence is driven at large scales and the magnetic
field is self-generated.
No hyperviscosity or hyper-resistivity is used.

\section{Equations}

Here we consider subsonic turbulence in an isothermal
electrically conducting
gas with constant sound speed $c_{\rm s}$ in a periodic box of size
$2\pi\times2\pi\times2\pi$.
The governing equations are
\EQ
{\DD\uu\over\DD t}=-c_{\rm s}^2\nab\ln\rho+{\JJ\times\BB\over\rho}
+\FF_{\rm visc}+\ff,
\label{dudt}
\EN
where $\DD/\DD t=\partial/\partial t+\uu\cdot\nab$ is the advective
derivative, $\JJ=\nab\times\BB/\mu_0$ the current density,
$\BB$ the magnetic field, $\mu_0$
the vacuum permeability,
\EQ
\FF_{\rm visc}=\nu\left(\nabla^2\uu+\onethird\nab\nab\cdot\uu
+2\SSSS\cdot\nab\ln\rho\right)
\EN
is the viscous force where
$\nu=\mbox{const}$ is the kinematic viscosity,
${\sf S}_{ij}=\frac{1}{2}(u_{i,j}+u_{j,i})-\frac{1}{3}\delta_{ij}u_{k,k}$
is the traceless rate of strain tensor, and $\ff$ is a
random forcing function (see below). The continuity equation is
written in terms of logarithmic density,
\EQ
{\DD\ln\rho\over\DD t}=-\nab\cdot\uu,
\EN
and the induction equation is solved in terms of the magnetic vector
potential $\AAA$, where $\BB=\nab\times\AAA$, and
\EQ
{\partial\AAA\over\partial t}=\uu\times\BB+\eta\nabla^2\AAA,
\label{dAdt}
\EN
and $\eta=\mbox{const}$ is the magnetic diffusivity (or resistivity); 
we choose $\eta=\nu$, i.e.~our magnetic Prandtl number is unity.
We adopt a forcing function $\ff$ of the form
\EQ
\ff(\xx,t)=\mbox{Re}\{N\ff_{\kk(t)}\exp[i\kk(t)\cdot\xx+i\phi(t)]\},
\EN
where $\xx=(x,y,z)$ is the position vector, and $-\pi\le\phi(t)<\pi$ is
a ($\delta$-correlated) random phase.
The normalization factor is
$N=f_0 c_{\rm s}(kc_{\rm s}/\delta t)^{1/2}$, with $f_0$ a
nondimensional forcing amplitude, $k=|\kk|$, and $\delta t$ the length of
the timestep;
we chose $f_0=0.02$ so that the maximum Mach number stays below about 0.5
(the rms Mach number is close to 0.12 in all runs.)
$\ff_{\kk}$ describes nonhelical transversal waves with
$|\ff_{\kk}|^2=1$ and
$\ff_{\kk}=\left(\kk\times\eee\right)/\sqrt{\kk^2-(\kk\cdot\eee)^2}$,
where $\eee$ is an arbitrary unit vector.
At each timestep we select randomly one of 20 possible wavevectors
in the range $1\leq |\kk|<2$ around the forcing wavenumber, $k_{\rm f}=1.5$.

The equations are solved using the same method as in \citet{B01}, but
here we employ a new cache and memory efficient code\footnote{
We use the Pencil Code which is a cache efficient
grid based high order code (sixth
order in space and third order in time) for solving the compressible
MHD equations; {\sf http://www.nordita.dk/data/brandenb/pencil-code}.}
using MPI (Message Passing Interface) library calls for communication between
processors.

\section{Results}

\begin{figure}[t!]\centering\includegraphics[width=0.45\textwidth]
{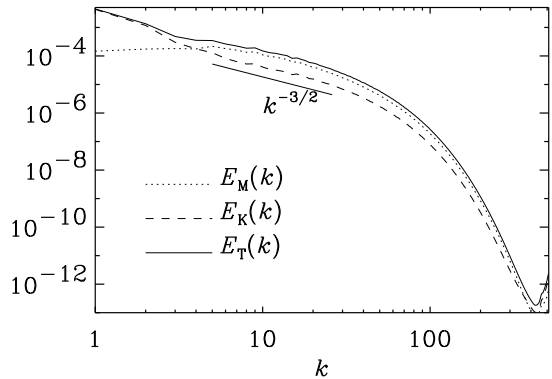}\caption{
Magnetic, kinetic and total energy spectra.
$1024^3$ meshpoints.
The Reynolds number is $u_{\rm rms}/(\nu k_{\rm f})\approx960$.
}\label{power1024a}\end{figure}

\begin{figure}[t!]\centering\includegraphics[width=0.45\textwidth]
{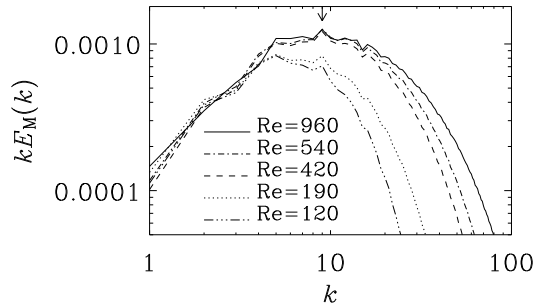}\caption{Spectral magnetic energy per logarithmic wavenumber interval
for runs with different 
Reynolds numbers. The arrow correspond 
to the peak ($k=9$) of the four largest runs.
}\label{mag0_k_comp}\end{figure}

In \Fig{power1024a} we plot magnetic, kinetic, and total energy spectra,
$E_{\rm M}(k)$, $E_{\rm K}(k)$, and $E_{\rm T}=E_{\rm M}+E_{\rm K}$,
respectively, for our largest resolution run with $1024^3$
meshpoints.\footnote{These spectra are, as usual, integrated
over shells in $k$ space and normalized such that
$\int E_{\rm K}\dd k={1\over2}\bra{\uu^2}$ and
$\int E_{\rm M}\dd k={1\over2}\bra{\BB^2}/\mu_0$.}
The magnetic energy displays a nearly flat spectrum in the range
$1\leq k\leq5$, peaks at $k\approx5$,
and begins to show an inertial range in $8\leq k\leq25$, followed by
a dissipative subrange over one decade.
In the inertial range $E_{\rm M}(k)/E_{\rm K}(k)$ is about 2.5.
A plot of magnetic energy per logarithmic wavenumber interval,
$k E_{\rm M}(k)$, shows that most of the magnetic energy comes from a
band of wavenumbers around $k=9$ -- independent of the Reynolds
number once the latter is above $\sim400$; see \Fig{mag0_k_comp}.

\begin{figure}[t!]\centering\includegraphics[width=0.45\textwidth]
{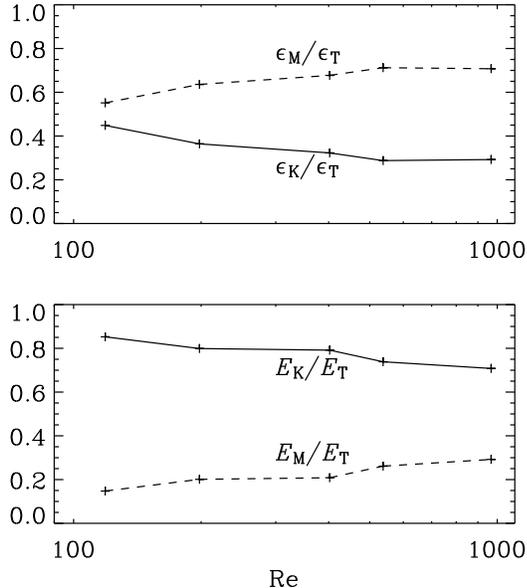}\caption{
Approximate convergence of
relative magnetic and kinetic energy dissipation rates (upper panel)
and the relative energies (lower panel)
as a function of Reynolds number.
}\label{Fenergy_dissipation_ratio}\end{figure}

The energy ratio in the inertial range is similar to the ratio of the
dissipation rates of magnetic and kinetic energies,
$\epsilon_{\rm M}/\epsilon_{\rm K}$, which is about 2.3;
see \Fig{Fenergy_dissipation_ratio}, which shows that magnetic and
kinetic energy dissipation rates approach 70\% and 30\%, respectively,
of the total dissipation rate, $\epsilon_{\rm T}=\epsilon_{\rm M}+\epsilon_{\rm K}$.
The convergence of relative dissipation rates is compatible with
Kolmogorov's concept of a constant, scale independent energy flux across
the spectrum, which seems to apply also separately for velocity and magnetic
fields.
This picture would be difficult to reconcile if the magnetic energy
were to always peak at the resistive scale.
We emphasize that the inertial range is not representative of the total
energy which, in turn, is dominated by small wavenumbers.
The ratio of total magnetic to kinetic energies is only 0.4 and
seems again to be asymptotically independent of Reynolds number
(\Fig{Fenergy_dissipation_ratio}).

\begin{figure}[t!]\centering\includegraphics[width=0.45\textwidth]
{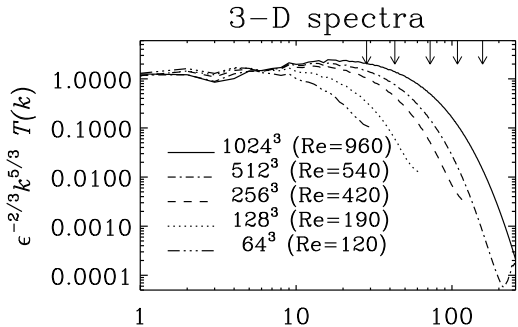}\caption{
Total energy spectra compensated by $\epsilon_{\rm T}^{-2/3}k^{5/3}$,
where $\epsilon_{\rm T}$ is the rate of total energy dissipation.
Spectra for different resolution are compared.
The dissipation cutoff wavenumber, $k_{\rm d}=(\epsilon_{\rm T}/\nu^3)^{1/4}$,
is indicated by short arrows at the top of the plot.
The Reynolds number, $\mbox{Re}=u_{\rm rms}/(\nu k_{\rm f})$, is given
in the legend.
}\label{kolmogorov3d}\end{figure}

By comparing runs at different resolution one can clearly see that
in the range $3\le k\le20$
the total energy spectrum is shallower than $k^{-5/3}$ (\Fig{kolmogorov3d}).
This could perhaps be due to the bottleneck effect that is known to
exist also in wind tunnel turbulence, where it
has been described by a weak $k^{-1}$ contribution \citep{SJ93}.
There are several reasons why such a bottleneck effect might occur.
First, recent studies by \citet{DHYB03} have shown that
the bottleneck effect is much stronger in shell integrated
three-dimensional spectra compared to just longitudinal or transversal
one-dimensional spectra
available in wind tunnel turbulence.
Second, the bottleneck effect may simply be stronger
for hydromagnetic turbulence
due to the dynamo effect that is expected to produce magnetic energy
preferentially at $k\approx k_{\rm f}R_{\rm c}^{1/2}\approx5...10$, where
$R_{\rm c}\approx30$ is the critical magnetic Reynolds number for
dynamo action \citep{Sub99}.
Third, numerical effects such as hyperdiffusion or other effects causing
departures from the physical $\nabla^2$ diffusion operator could cause
an artificial bottleneck effect \citep{BM00,BSC98}, which may also explain
the extended $k^{-1}$ range in compressible nonmagnetic simulations
using the piecewise parabolic method \citep{PWP98}.

We believe that numerical effects do not cause such artifacts in our simulations,
because we use the physical $\nabla^2$ diffusion operator and our
discretization is accurate to sixth order in space and third order in time.
We have compared two runs with identical values of $\nu$
and $\eta$, one with $256^3$ meshpoints and the other one with $512^3$,
and found the same spectra \citep{HBD03a}.
We have also compared with runs using double precision and found the
spectra to be the same.
The runs with up to $512^3$ meshpoints have run for up to
80 turnover times, $(u_{\rm rms}k_{\rm f})^{-1}$, which should be long
enough to eliminate transients.
The run with $1024^3$ meshpoints has only run for 5 turnover times,
but the results are otherwise in qualitative agreement with those
of the $512^3$ run.

By comparing with nonmagnetic runs at the same resolution ($1024^3$
meshpoints) we have found that in hydromagnetic turbulence the bottleneck
effect is about equally strong.
This leaves us with the possibility that the bottleneck effect is real,
both for hydrodynamic and for hydromagnetic turbulence, but that it is
simply more pronounced in fully three-dimensional spectra.
It is therefore still possible that at larger Reynolds numbers the
true inertial range spectrum will have a $k^{-5/3}$ behavior both for
kinetic and magnetic energies \citep{GS95}.

Alternatively, a $k^{-3/2}$ spectrum might be readily explicable in terms
of the Iroshnikov-Kraichnan phenomenology \citep{Iro63,Kra65}.
It does of course ignore local anisotropy, but more importantly, it predicts
that the fourth order structure function, $S_4(\ell)$,
scales linearly in the inertial range,
$\zeta_4 \equiv d\ln S_4/d\ln\ell = 1$; see \citet{Bis93}.
To assess this possibility, we calculate the double-logarithmic derivative,
$\zeta_p=\dd\ln S_p/\dd\ln\ell$, of the
unsigned structure function, $S_p=\bra{|\zzz^\pm|^p}$,
where $\zzz^\pm=\uu\pm\BB/\sqrt{\mu_0\rho}$ are the Elsasser variables.
The scalings of $\zzz^+$ and $\zzz^-$ turn out to be similar, so we take
the average value.

\begin{figure}[t!]\centering\includegraphics[width=0.45\textwidth]
{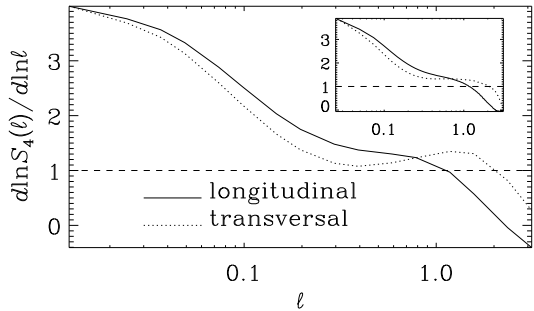}\caption{
Fourth order structure function for runs with $512^3$ meshpoints.
Clearly, $S_4(\ell)$ is {\it not} compatible with linear scaling.
The inset gives the result for $256^3$ meshpoints.
The scaling for transversal structure functions (dotted lines) tends
to be better than for the longitudinal ones (solid lines).
The statistics for the $256^3$ runs is somewhat better
than for the shorter $512^3$ runs.
}\label{fourth_moment}\end{figure}

\begin{figure}[t!]\centering\includegraphics[width=0.45\textwidth]
{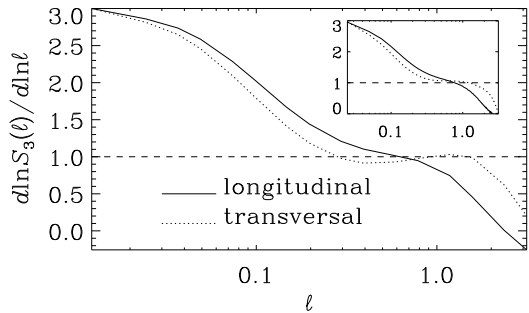}\caption{
Same as \Fig{fourth_moment}, but for $S_3(\ell)$.
Note that $S_3(\ell)$ is compatible with linear scaling,
i.e.\ $\zeta_3=1$ in the inertial range.
}\label{third_moment}\end{figure}

Figure~\ref{fourth_moment} shows that
our data are {\it inconsistent} with Iroshnikov-Kraichnan scaling
($\zeta_4\approx1.3$, rather than $1$). 
Instead, our data are consistent with $\zeta_3=1$ in the inertial
range (\Fig{third_moment}).
Note that linear inertial range scaling of $S_3(\ell)$, i.e.\ $\zeta_3=1$,
is an exact result for hydrodynamic turbulence \citep{Fri95}.
Using the extended self-similarity hypothesis
\citep{Benzi93} we plot $S_p(\ell)$ against $S_3(\ell)$ to
obtain a more accurate determination of $\zeta_p$ for $p\neq3$.
Our results are compatible with \citet{SL94}, and hence
also with earlier forced simulations with an external field \citep{CLV02a}.
In particular, $\zeta_2+1$ is the negative slope of the total energy
spectrum: we find $\zeta_2=0.70$, which is again in agreement with
Kolmogorov scaling.
This agrees also with simulations of decaying helical
turbulence \citep{BM00}, except that there the higher moments have
been found to be smaller.

\section{Conclusions}

For nonhelically forced hydromagnetic turbulence a resolutions of $1024^3$
is necessary in order to begin to establish inertial range scaling.
Nevertheless,
the energy spectra show what we interpret as a strong bottleneck effect.
This effect is particularly strong in the shell-integrated spectra.
Throughout the inertial range, however, the magnetic energy exceeds
the kinetic energy by a factor of about 2.5.
Both kinetic and magnetic energies are dominated by the spectral values
at the beginning of the inertial range and independent of magnetic resistivity.
The values $\zeta_2=0.70$ and $\zeta_3=1.0$ strongly favor an
asymptotic $k^{-5/3}$ spectrum.

Our results do not support recent claims that the magnetic energy spectrum
peaks at the resistive scale \citep{MC01,MB02,Scheko02a,Scheko02b}.
However, closer inspection of Fig.~3 of \citet{MC01} or Fig.~1 of
\citet{MB02} reveals that also in their cases the magnetic energy
peaks at about 5, in agreement with our results.
This is comparable to the value $k_{\rm f}R_{\rm c}^{1/2}\approx8$
expected based on a nonlinear closure model \citep{Sub99}.
Here, $R_{\rm c}\approx30$ is the critical magnetic Reynolds number for
dynamo action \citep{HBD03a}, and $k_{\rm f}\approx1.5$.
For $k\geq8$ we find the spectrum to be Kolmogorov-like.

In the interstellar medium, the magnetic Prandtl number is very large
\citep{KA92}.
This is also the regime in which the claims regarding
the peak at the resistive scale are thought to apply best.
On the other hand,
preliminary results suggest that even for a magnetic Prandtl number
between 5 and 30 the magnetic energy spectrum still peaks at $k\approx5$
\citep{HBD03a}.
However, the spectrum shows now a possible $k^{-1}$ tail near the viscous
dissipation range where ohmic dissipation is still weak \citep{CLV02b,HBD03a}.
In the interstellar medium, even though the magnetic Prandtl number
is very large, the Reynolds number is also very large, so the viscous
cutoff scale is still small ($\sim10^{16}\cm=0.003\pc$).
We expect that in the range $0.01\dots10\pc$ kinetic and magnetic
energy spectra are parallel and show $k^{-5/3}$ scaling, but currently
the numerical resolution is still insufficient to demonstrate this.

\acknowledgments
We thank Jason Maron, {\AA}ke Nordlund and Kandaswamy Subramanian
for their comments on the paper.
Use of the parallel computers in Trondheim (Gridur), Odense (Horseshoe)
and Leicester (Ukaff) is acknowledged.

%r e f

\end{document}